\documentclass[twocolumn,showpacs,preprintnumbers,amsmath,amssymb]{revtex4}

\usepackage{epsfig}
\usepackage{amsmath}
\usepackage{epsfig}
\usepackage{psfrag}
\usepackage{amsfonts}
\usepackage{graphicx}
\usepackage{dcolumn}
\usepackage{bm}
\usepackage{lineno}

\usepackage{graphicx}
\usepackage{dcolumn}
\usepackage{bm}
\newcommand{\vect}{\left ( \begin{array}{c}}
\newcommand{\evect}{\end{array} \right )}

\usepackage{epsfig}
\usepackage{psfrag}
\usepackage{graphicx}
\usepackage{dcolumn}
\usepackage{bm}

\begin{document}

\title{The influence of magnetic field on the pion superfluidity and phase structure in the NJL model }
\author{Xiaohan Kang}
\affiliation{Key Laboratory of Quark and Lepton Physics (MOE) and Institute of Particle Physics,
Central China Normal University, Wuhan 430079, China}
\author{Meng Jin}
\affiliation{Key Laboratory of Quark and Lepton Physics (MOE) and Institute of Particle Physics,
Central China Normal University, Wuhan 430079, China}
\author{Juan Xiong}
\affiliation{Department of Applied Physics, College of Sciences,
Huanzhong Agriculture University, Wuhan 430070,China}
\author{Jiarong Li}
\affiliation{Key Laboratory of Quark and Lepton Physics (MOE) and Institute of Particle Physics,
Central China Normal University, Wuhan 430079, China}

\begin{abstract}
The influence of the magnetic field on the pion superfluidity and
the phase structure is analyzed in the framework of the two-flavor
Nambu--Jona-Lasinio(NJL) model. To do this, we first derive the
thermodynamic potential from the Lagrangian density of the NJL
model in the mean field approximation. Using this thermodynamic
potential, we get the gap equation of the chiral condensate and
the pion condensate. The effect of external magnetic field on the
pion condensate is not simple promotion or suppression, which we
will discuss in detail in the paper. It is shown that the
tricritical point on the pion superfluidity phase transition line
moves to the space with smaller isospin chemical potential and
higher temperature when the external magnetic field becomes
stronger. The influence of external magnetic field on the chiral
condensate is also studied.
\end{abstract}

\date{\today}

\maketitle

\section {Introduction}
\label{s1}
Quantum Chromodynamics (QCD) is the gauge theory of strong
interaction. Recently, the QCD phase diagram including chiral
symmetry breaking
(restoration)~\cite{ynambu,rchwa,spklevansky,masakawa,jhufner,pzhuang,dpmenezes,ajmizher,fpreis},
quark confinement (deconfinement)~\cite{cratti,sroessner}, color
superconductivity~\cite{malford,rrapp,mhuang,debert1,dhrischke,mbuballa}
were investigated with finite temperature and finite chemical
potential. Furthermore, the studies are extended to finite isospin
chemical
potential~\cite{dtson,mcbirse,jbkogut,abmigdal,dtoublan,abarducci,lhe,cmu}.
It is found that at a critical isospin chemical potential, which
is about the pion mass in vacuum, the phase transition from normal
phase to pion superfluidity phase will happen. The understanding
of the properties of pion superfluidity and phase structure under
the extreme conditions are important in conducting research of
the physics of the compact objects and relativistic heavy ion
collisions.

It is reported that very strong magnetic field may be generated in
the heavy-ion collision~\cite{ivsely,dekharzeev1,dekharzeev2}. The
magnetic field magnitudes were estimated to be 5.3$m_\pi^2/e$ at
the Relativistic Heavy-Ion Collision(RHIC) and 6$m_\pi^2/e$ at the
Large Hadron Collision(LHC), and even higher~\cite{vskokov}. Such
magnetic field can also exist in magnetars~\cite{rduncan} and in
the early universe~\cite{msturner,bratra,lmwidrow} despite the
origin of such strong field is not clear yet. Above phenomena lead
us to think deeply about the influence of the magnetic field on
the QCD phase diagram~\cite{gnferrari}. Recently, much work has
been done about the influence of the magnetic field on the
properties of quark matter~\cite{dekharzeev3}, chiral
transition~\cite{joandersen} and color
superconductivity~\cite{jlnoronha, lpaulucci, ejferrer}.

Normally, QCD is widely accepted as the correct theory describing
strongly interacting matter at high temperature and high density.
The low energy QCD vacuum is hard to be fully understood by using
of perturbative methods, because the characteristics of chiral
symmetry breaking and color confinement have a non-perturbative
origin. To solve the problem, two approaches are introduced, one
is lattice QCD simulations~\cite{debert2} and the other is the
effective model method.

The Nambu--Jona-Lasinio (NJL)
model~\cite{ynambu,debert3,mkvolkov,debert4,debert5} is used as a
simple and practical chiral model, which satisfies the basic
mechanism of spontaneous breaking of chiral symmetry and key
features of QCD at finite temperature and chemical potential. One
of the basic properties of this model is that it includes a gap
equation which connect the chiral condensate to the dynamical
quark mass. It is well known that the hadronic mass spectra and
the static properties of meson, especially the chiral symmetry
spontaneous breaking, can be obtained through the mean field
approximation (RPA) of meson in the NJL model. This model and its
extended version (PNJL, EPNJL, etc.) are also widely used to study
the properties of deconfinement phase, the color superconductivity
phase, and the pion superfluidity phase and the related phase
transitions under extreme conditions. Recently, the strong
magnetic field effect on the properties of quark
matter~\cite{dpmenezes2,ssavancini,schak}, and the phase
transitions,
 including chiral restoration transition, deconfinement transition~\cite{rgatto, kkashiwa} and color superconductivity transition~\cite{ejferrer, tmandal}, has been investigated in the NJL-type models by many groups.

In this paper, we will mainly focus on the effect of external
magnetic fields on the pion condensate and the pion superfluidity
phase structure at finite temperature and finite isospin chemical
potential. The influence of the magnetic field on the chiral
condensate is also studied.

This paper is organized as follows. In Sec. II, we will give a
simple introduction of our model and calculate the thermodynamics
potential within mean field approximation. Sec. III is devoted to
the numerical results of the quark pair condensation and the phase
diagram in the $T-\mu_I-eB$ space. We summarize and conclude our
job in Sec. IV.

\section {the model}
\label{s2}
The Lagrangian density of two-flavor Nambu-Jona-Lasinio (NJL)
model is defined as:

\begin{eqnarray}
\mathcal {L}=\bar{\psi}(i\gamma_\mu
D^\mu-\hat{m}_0)\psi+G[(\bar{\psi}\psi)^2+(\bar{\psi}i\gamma_5\vec{\tau}\psi)^2],
\end{eqnarray}

where $\psi=(\psi_u,\psi_d)^T$ is the quark field, $\hat{m_0}$ =
$diag$ $(m_u, m_d)$ is the current quark mass matrix with
$m_u=m_d\equiv m_0$ (the isospin symmetry).
$\mathcal{D}^\mu=\partial^\mu+ieA_\mu$ is the covariant
derivative, $A^\mu=\delta^\mu_0A^0$ and $A^0=-iA_4$ represents the
gauge field. $G$ is the four-quark coupling constant with
dimension $GeV^{-2}$. The Pauli matrices $\tau_i$ $(i=1, 2, 3)$
are defined in isospin space.

With scalar and pseudoscalar interactions corresponding to
$\sigma$ and $\pi$ excitation, the lagrangian density has the
symmetry of $U_B(1)\bigotimes SU_I(2)\bigotimes SU_A(2)$,
corresponding to baryon number symmetry, isospin symmetry and
chiral symmetry, respectively. The chiral symmetry $SU_A(2)$
breaks down to $U_A(1)$ global symmetry which is associated with
the chiral condensation of the $\sigma$ meson.
\begin{eqnarray}
\sigma=<\bar{\psi}\psi>=\sigma_u+\sigma_d,\\
\sigma_u=<\bar{u}u>, \sigma_d=<\bar{d}d>.
\end{eqnarray}
The isospin symmetry $SU_I(2)$ breaks down to $U_I(1)$ global
symmetry with the generator $I_3$ which is related to the
condensate of charged pions, $\pi^+$ and $\pi^-$,
\begin{eqnarray}
\pi^+=<\bar{\psi}i\gamma_5\tau_+\psi>=\sqrt{2}<\bar{d}i\gamma_5u>,\\
\pi^-=<\bar{\psi}i\gamma_5\tau_-\psi>=\sqrt{2}<\bar{u}i\gamma_5d>.
\end{eqnarray}
With $\tau_\pm=(\tau_1\pm\tau_2)/\sqrt{2}$.  At extremely high
$\mu_I>0$ the condensate of $u$ and anti-$d$ quark is favored. At
extremely high $\mu_I<0$ the condense of $d$ and anti-$u$ quark is
favored. The system is in the global thermal equilibrium,
$\pi^+=\pi^-$, as is assumed in this paper, the whole
superfluidity is electric charge neutral.

By means of the mean field approximation, the thermodynamic
potential of two flavor NJL model at finite isospin chemical
potential, finite temperature and strong magnetic field is given
as
\begin{eqnarray}
\Omega&=&G(\sigma^2+\pi^2)-2N_c\sum_{f=u,d}\sum_{\kappa}\alpha_\kappa
\nonumber\\&&\int^{+\infty}_{-\infty}\frac{dp_z}{2\pi}\frac{|Q_f
eB|}{2\pi}[\omega^\pi_f+2Tln(1+e^{-\beta\omega^\pi_f})],~~
~\label{eq:OB}
\end{eqnarray}
with $\omega^\pi_f$ on behalf of $\omega^\pi_u$ and $\omega^\pi_d$, the quasi-particle energy of $u$ and $d$ quark.
$Q_f$ means the electric charge of $Q_u$ and $Q_d$. $\kappa$ is non-negative integer which denotes the Landau level
and $\alpha_\kappa=2-\delta_{\kappa 0}$ is the corresponding degeneracy. In the equation (\ref{eq:OB}),
\begin{equation}
 \begin{split}
\omega^\pi_f=\sqrt{(\omega_f\pm\mu_I)^2+4G^2\pi^2},~\\
\omega_f=\sqrt{p_z^2+2|Q_feB|\kappa+M^2},~\\
 \end{split}
\end{equation}
and
\begin{eqnarray}
M=m_0-2G\sigma.~
\end{eqnarray}

If we use the replacement for the momentum integral and the quark
energy,
\begin{eqnarray}
2\int\frac{d^3\bm p}{(2\pi)^3}\leftrightarrow\frac{|Q_f eB|}{2\pi}\sum_{\kappa}\alpha_\kappa\int\frac{dp_z}{2\pi},\\
\sqrt{\bm p^2+M^2}\leftrightarrow\sqrt{p_z^2+2|Q_f eB|\kappa+M^2},
\end{eqnarray}
the thermodynamic potential $\Omega$ above will coincide with that
in the zero magnetic field case \cite{lhe}. For the upper limit of
$p_z$ integral, we use the hard cut
$\sqrt{\Lambda^2-2\kappa|eB|}$. Then the upper limit of $\kappa$
sum is $\kappa_{max}=Int[\frac{\Lambda^2}{2|Q_feB|}]$, for
$Q_u=+\frac{2}{3}e$ and $Q_d=-\frac{1}{3}e$.

The gap equation of the mean field $\sigma$ and $\pi$ are derived
from
\begin{eqnarray}
\frac{\partial\Omega}{\partial\sigma}=0,\frac{\partial\Omega}{\partial\pi}=0.
\end{eqnarray}
When there exist multi-roots, only the solution which satisfies
the minimum condition is physical. The gap equation of $\sigma$ is \\
\begin{eqnarray}
\sigma+2N_c\int^{+\infty}_{-\infty}\frac{dp_z}{2\pi}\sum_\kappa\alpha_\kappa\frac{M|eB|}{2\pi}\{Q_u[1-2f(\omega^\pi_u)]\nonumber
\\\frac{\omega_u+\mu_I}{\omega^\pi_u\omega_u}
+Q_d[1-2f(\omega^\pi_d)]\frac{\omega_d-\mu_I}{\omega^\pi_d\omega_d}\}=0.~~
\end{eqnarray}
The gap equation of $\pi$ is
\begin{eqnarray}
1-4N_cG\sum_{f=u,d}\sum_\kappa\alpha_\kappa\int^{+\infty}_{-\infty}\frac{dp_z}{2\pi}\nonumber\\\frac{|eB|}{2\pi}\frac{Q_f[1
-2f(\omega^\pi_f)]}{\omega^\pi_f}=0.
\end{eqnarray}
$\pi=0$ is the trival solution of this equation, which has been
ignored here. The Fermi function is $f(x)=\frac{1}{(1+e^{\beta
x})}$. We choose the following values as numerical analysis
parameters: $m_0=5MeV$, a three-dimensional momentum cut-off
$\Lambda=650.7MeV(T=0)$, $G=5.01GeV^{-2}$, $m_\pi=139MeV$ and
$|<\bar{\psi}_u\psi_u>|^{1/3}=-250MeV$.
\section {Numerical results}
\label{s3}
\subsection {the pion and chiral condensate}
\begin{figure}

\begin{center}
\includegraphics[width=7cm]{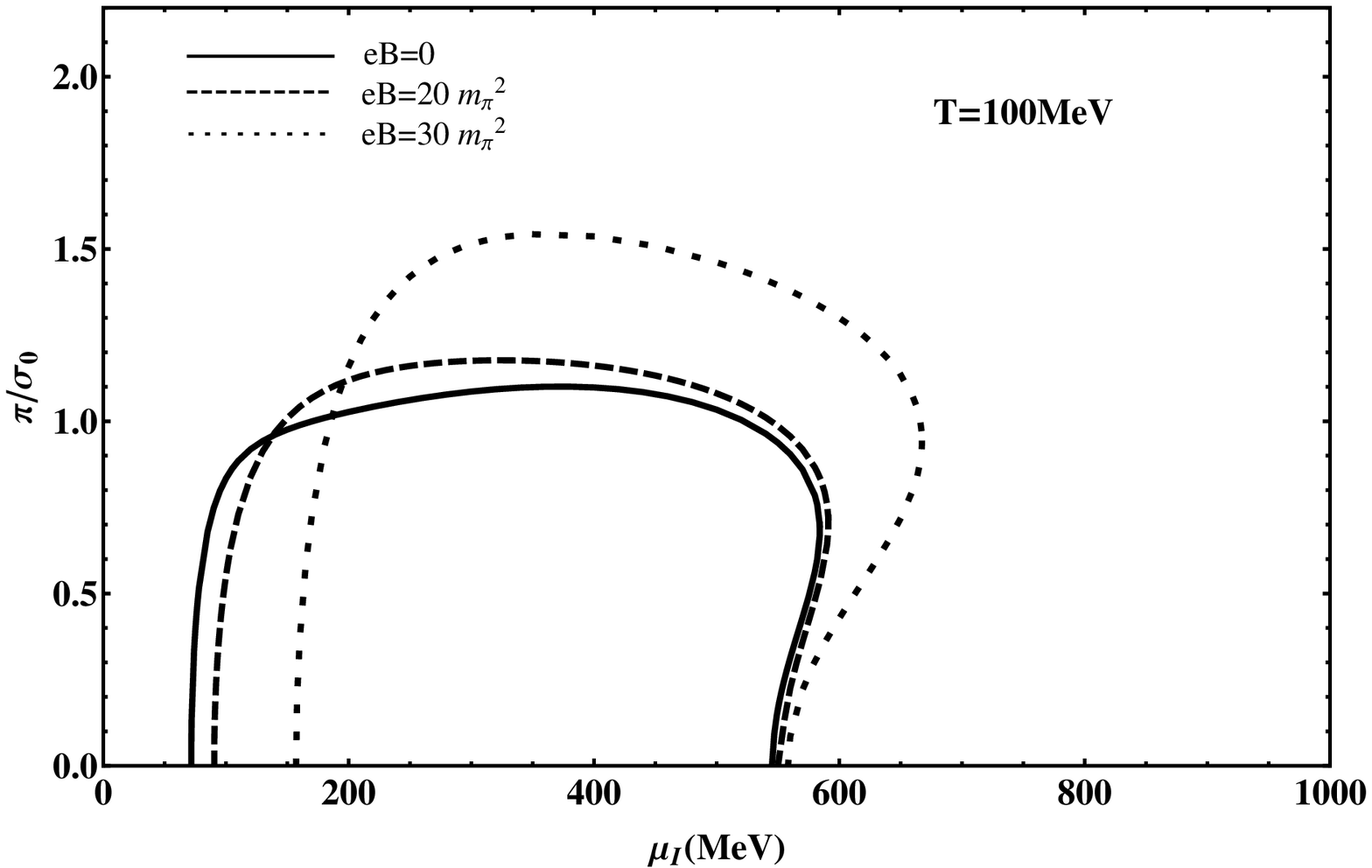} \\
\includegraphics[width=7cm]{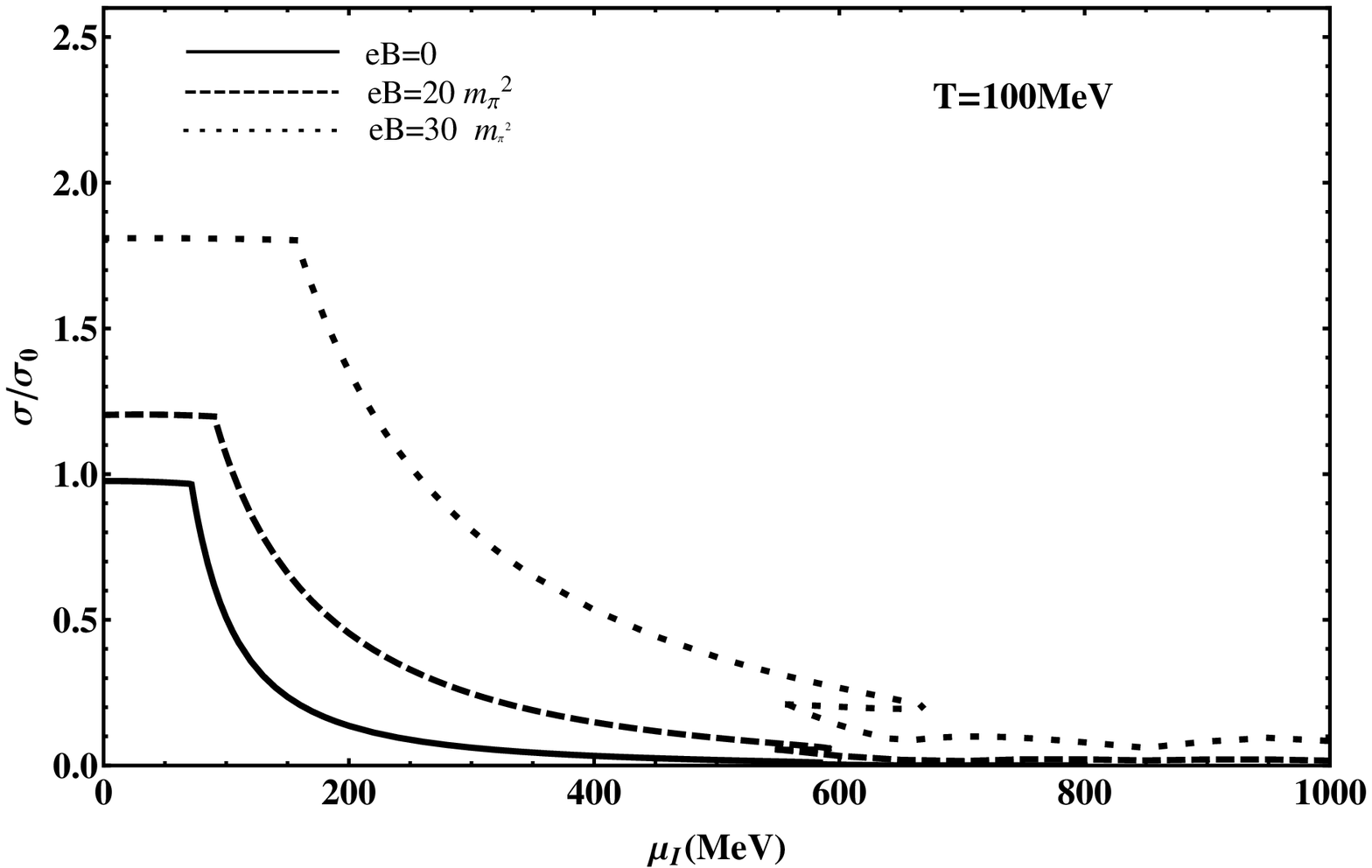} \\
\end{center}

\caption{{\em Upper panel:} Pion condensate $\pi$, shown as a
function of the isospin chemical potential $\mu_I$ with respect to
various magnetic field magnitudes. {\em Lower panel:} Chiral
condensate $\sigma$, shown as a function of the isospin chemical
potential $\mu_I$ with respect to various magnetic field
magnitudes.} \label{T100}
\end{figure}

In Fig. \ref{T100} we plot the behavior of the chiral condensate
$\sigma$ and the pion condensate $\pi$, measured in units of the
chiral condensate in vacuum $\sigma_0$, with varing isospin
chemical potential $\mu_I$ at fixed temperature $T=100 MeV$ for
various magnitude of the magnetic field, $eB=0$, $20m_\pi^2$, and
$30m_\pi^2$.

In the upper panel, for $eB=0$, we reproduced the result of
Ref.~\cite{lhe}. The $\pi$ superfluidity begin at critical isospin
chemical potential $\mu_I^c=m_\pi/2\approx70MeV$, then the pion
condensate increases when $\mu_I$ becomes larger, which says that
the isospin chemical potential enhance the pion condensate
\footnote{we notice that when the $\mu_I$ is large enough, the
pion condensation finally decrease to zero. The physics in this
region may be
 related to the model we used.}.

When the external magnetic field is included, such as $eB=20
m_\pi^2$ and $30 m_\pi^2$, the critical $\mu_I$ for occurrence of
pion condensate increases with the increasing magnetic fields,
indicating the magnetic field suppresses the formation of the pion
superfluidity. For different magnetic fields, the pion condensate
gaps remain the similar shape as a function of $\mu_I$, the
difference is that the maximum of gap is bigger with larger $eB$.
For example, the maximum of $\pi$ gap is $1.1 \sigma_0$ when
$eB=0$, but it reaches to $1.5 \sigma_0$ when $eB=30m_\pi^2$. For
$\mu_I > 550 MeV$, there exist three solutions for the $\pi$
condensation corresponding to a given isospin chemical potential
(one of them is zero). Such feature of the order parameter
normally indicates that the phase transition is of first order.

In the lower panel, for $eB=0$, the chiral condensation also
coincides with the result in Ref.~\cite{lhe}. The chiral
condensate decreases with increasing $\mu_I$, which reflects the
restoration of the chiral symmetry. When the external magnetic
fields are taken into account, such as $eB=20m_\pi^2$ and $30
m_\pi^2$, the $\sigma$ condensate is enhanced with increasing $B$,
this is the so called chiral catalytic effect of the magnetic
field~\cite{kgklimenko,vpgusynin}. The restoration of chiral
symmetry still happen with the increasing $\mu_I$. From the change
tendency of the $\sigma$ condensate we can roughly extract the
critical isospin chemical potential $\mu_I^c$. The bigger the
magnetic field, the bigger the $\mu_I^c$. This implies that the
external magnetic field hinders the restoration of chiral
symmetry.

\begin{figure}

\begin{center}
\includegraphics[width=7cm]{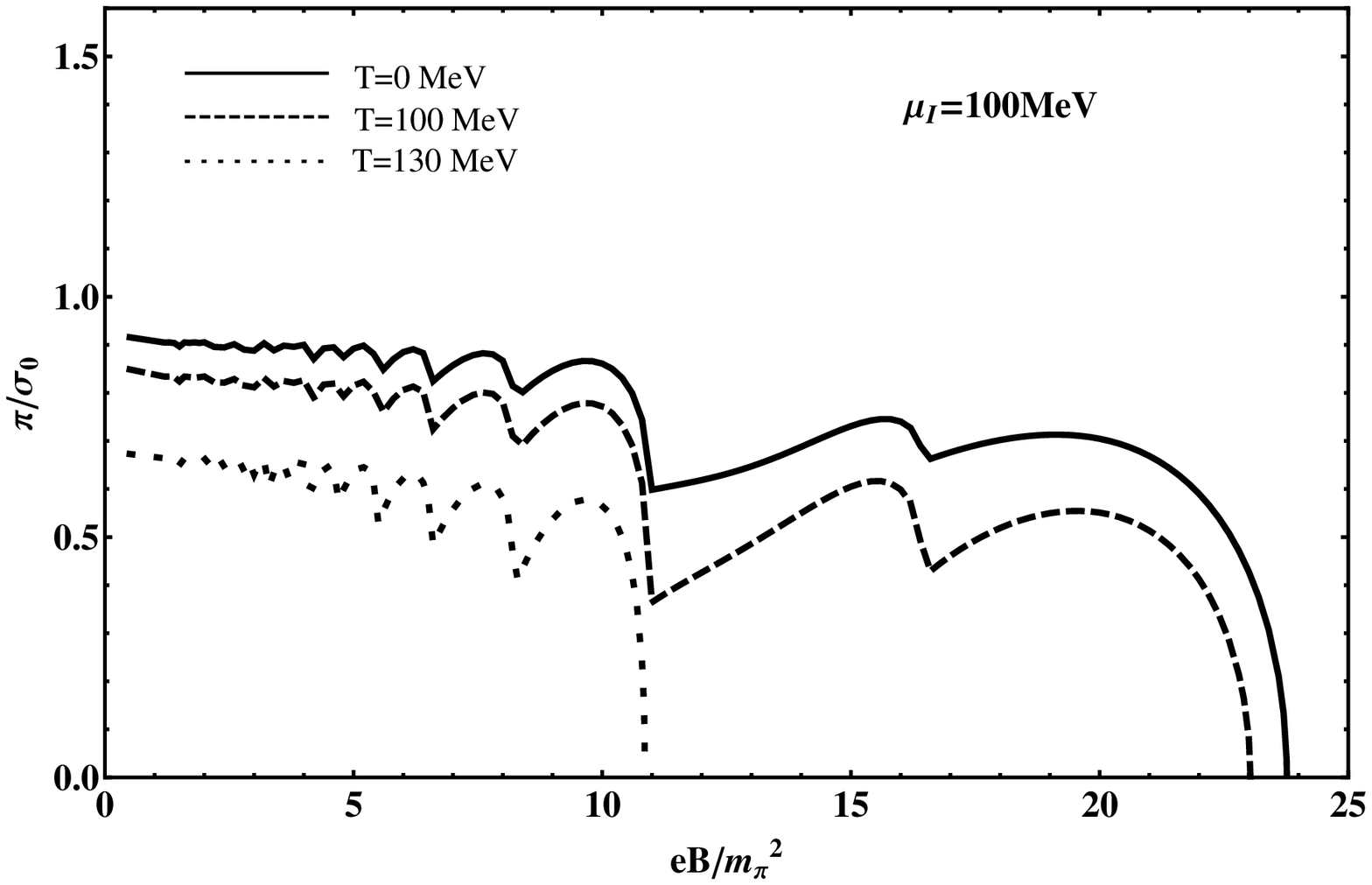} \\
\includegraphics[width=7cm]{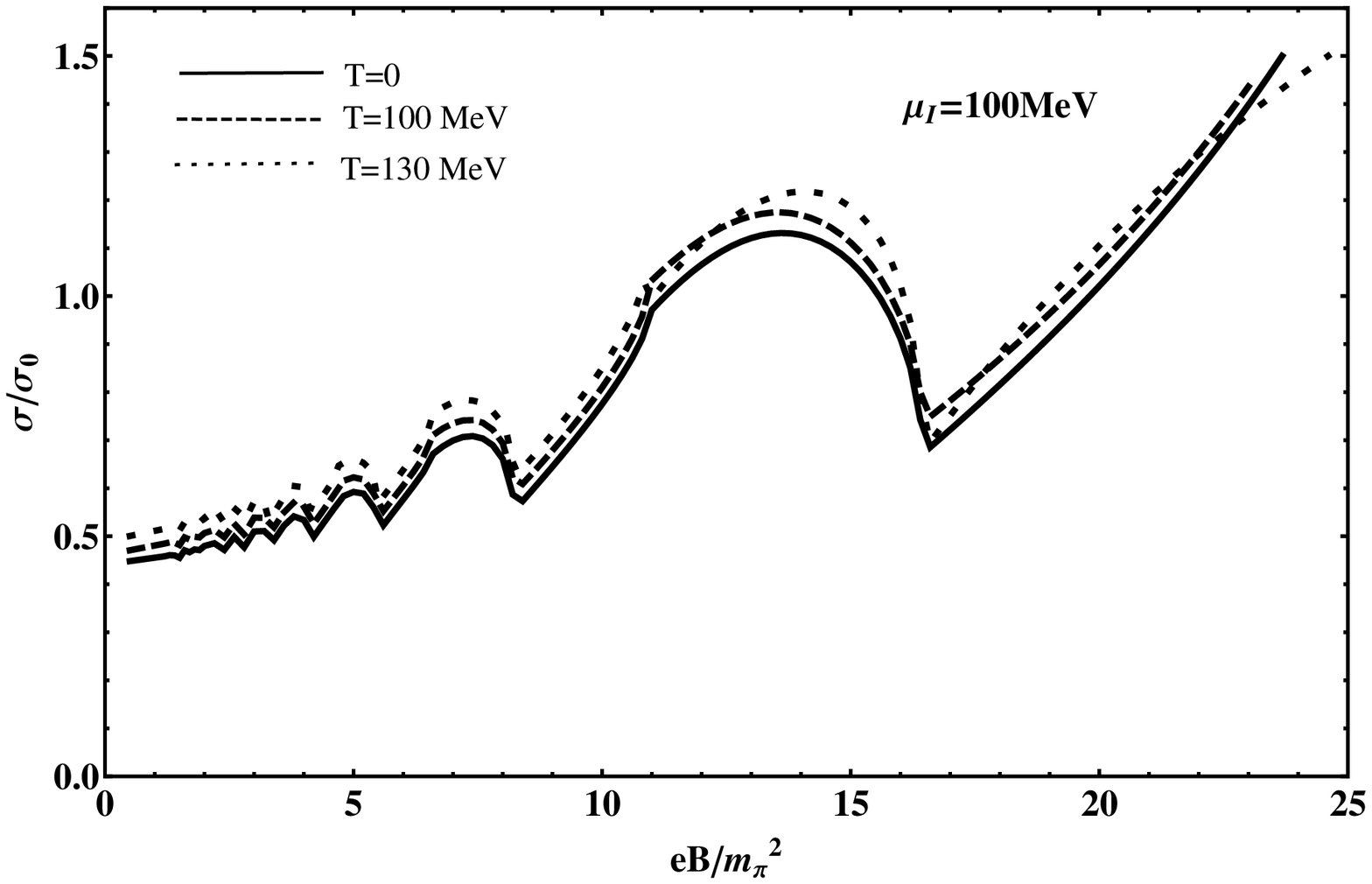} \\
\end{center}

\caption{{\em Upper panel:} Pion condensate $\pi$, shown as a
function of the magnetic field strength $eB$ with respect to
various temperature magnitudes. {\em Lower panel:} Chiral
condensate $\sigma$, shown as a function of the magnetic field
strength $eB$ with respect to various temperature values.}
\label{ui100}
\end{figure}

In Fig. \ref{ui100} we show the behavior of the chiral condensate
$\sigma$ and pion condensate $\pi$, with varing magnetic field
strength at fixed isospin chemical potential $\mu_I=100MeV$ for
several values of temperature, $T=0$, $100MeV$ and $130MeV$.

In the upper panel, the $\pi/\sigma_0$ almost remains stable with
small $eB$. When $eB$ increases to certain value, the $\pi$
condensation starts to decrease oscillately, and then disappears
when the external magnetic field is strong enough. This implies
that $\pi$ condensate is suppressed with increasing $eB$. From the
figure we can see the higher the temperature, the narrower range
of the magnetic field domain of $\pi\neq0$ and the smaller the
critical $eB$ for pion vanishing. That is to say the pion
condensate is suppressed by the temperature. An interesting
phenomenon is that the pion condensate exhibits an oscillation
behavior as a function of $eB$. From our calculation, the
oscillating behavior is only related to the external magnetic
field but not the temperature or the isospin chemical potential,
it is also verified in the lower panel.

In the lower panel, with increasing $eB$, the chiral condensate
increases oscillately. That is coincide with the chiral catalytic
effect. The oscillation is obvious even when the temperature is
very high, such phenomenon is the so-called Alfven-de Haas
oscillation. The oscillating behavior related to the external
magnetic field is similar with the $\pi$. But the oscillating
behavior vanishes when $eB>16.5 m_\pi^2$ and the $\sigma$
condensate increases monotonously with increasing $eB$. The reason
is that when the $eB$ is higher enough, the Lowest Landau Level
(LLL) will chiefly contribute to the properties of the system and
then the oscillation disappears. The influence of temperature on
chiral condensate is not such apparent as the pion condensation,
the curves of $\sigma/\sigma_0$ coincide approximately for the
three values of temperature we chose.

We noticed in some $eB$ regions, for example, between
$eB=13m_\pi^2\sim17m_\pi^2$, the chiral condensate will decrease
as the increasing $eB$. The decrease is a consequence of the
oscillating behavior of the $\sigma$ condensate, which is
different from the inverse chiral magnetic catalysis effect.

\begin{figure}

\begin{center}
\includegraphics[width=7cm]{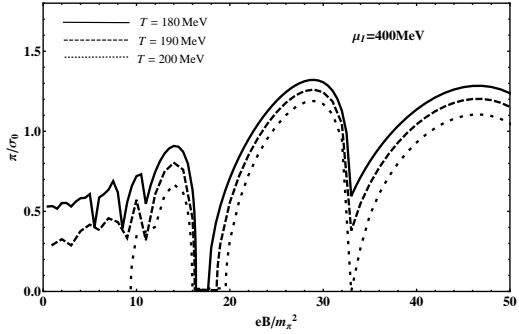} \\
\end{center}

\caption{The Pion condensate $\pi$, shown as a function of the
magnetic field strength B with respect to various temperature
magnitude at fixed isospin chemical potential $\mu_I=400MeV$.}
\label{ui400}
\end{figure}

In Fig. 3, we show the pion condensation varies with $eB$ for
$\mu_I=400MeV$. Now the pion condensation increase oscillately
with $eB$, impling the magnetic field can also enhance the pion
condensate. This contradicts the behavior for the case of $\mu=100
MeV$ in Fig. \ref{ui100}. Normally the isospin chemical potential
can promote the pion condensate while the temperature depress it.
However, the effect of magnetic field is complicated. In some case
it depresses the $\pi$ condensation and in the other case it
reinforce the $\pi$ condensation.  For some given temperature and
isospin chemical potential, such as $T>195MeV$, $\mu_I=400MeV$,
the pion condensate does not exist at $eB=0$, and then it occurs
as the $eB$ increases, showing that the external magnetic field
can promote the formation of the pion superfluidity. This
phenomenon is the result of the coupled influence of $\mu_I$ and
$eB$. When we increase $eB$ higher enough, the $\pi$ condensate
start to decrease and will disappear at some value of $eB$ (it is
not shown in the Figure), that is similar with the pion condensate
in Fig.2.

\begin{figure}

\begin{center}
\includegraphics[width=7cm]{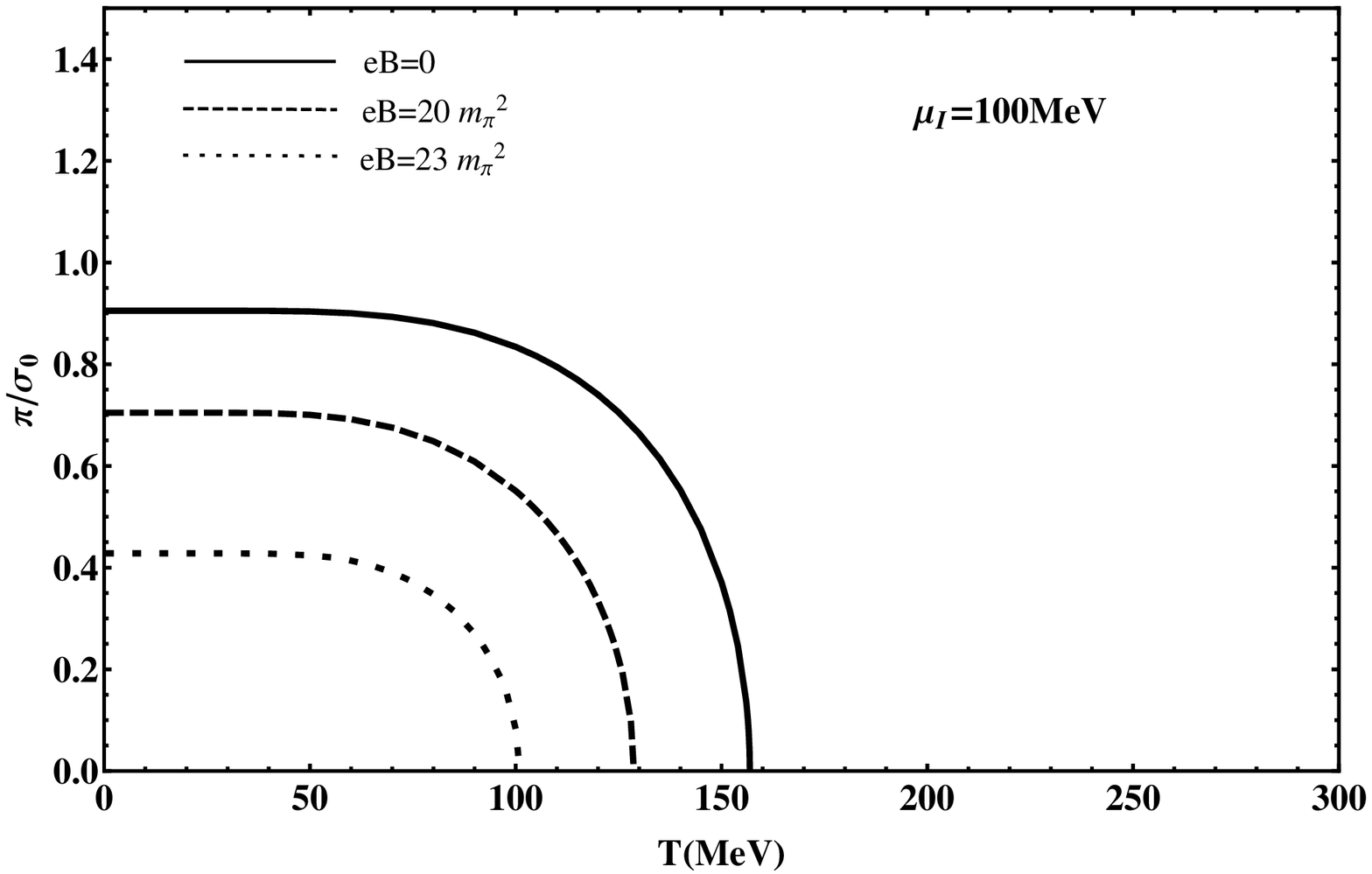} \\
\includegraphics[width=7cm]{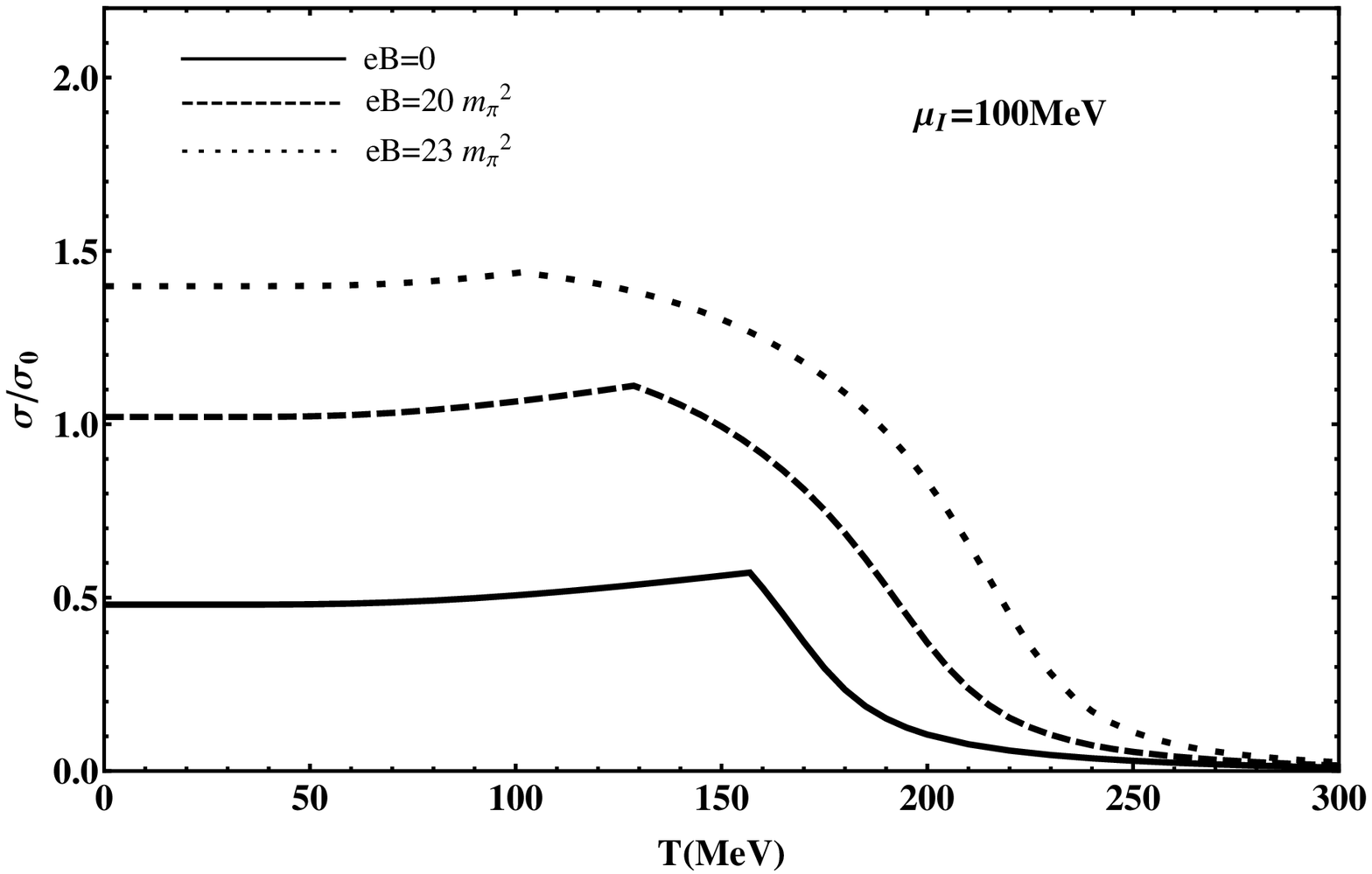} \\
\end{center}

\caption{{\em Upper panel:} Pion condensate $\pi$, shown as a
function of the temperature with respect to various magnetic field
magnitudes. {\em Lower panel:} Chiral condensate $\sigma$, shown
as a function of the temperature with respect to various magnetic
field magnitudes.} \label{uui100}
\end{figure}

In Fig. \ref{uui100} we show the behavior of the pion condensate
$\pi$ and the chiral condensate $\sigma$, with varing $T$ for
isospin chemical potential $\mu_I=100MeV$ but for various magnetic
field magnitudes, $eB=0$, $20m_\pi^2$ and $23m_\pi^2$,
respectively.

In the upper panel, for $eB=0$ the $\pi$ condensate decreases
monotonously as the temperature increases. When T increases to the
critical value, the pion condensate disappears. This implies that
the temperature depress the pion condensate, as is shown similarly
in Fig. 2 and Fig. 3. Taking the magnetic field into
consideration, the greater the $eB$, the narrower range of
temperature domain of the $\pi\neq 0$ and the lower the critical T
for $\pi$ superfluidity disappearing, reflecting that $\pi$
condensate is suppressed by the strong magnetic field for
$\mu_I=100MeV$. This result is self-consistent with what we have
obtained in Fig. 2.

In the lower panel, the $\sigma$ condensate rises slowly with
increasing temperature. The above phenomenon only happens in the
pion superfluidity phase.In the high $T$ region the $\pi=0$, the
$\sigma$ condensation decreases monotonously as the $T$ increases,
meaning the chiral symmetry will restore in this region. When
magnetic field is included, the varying behaviors of the chiral
condensation are similar, but the chiral condensate increases with
$eB$. That is to say that the magnetic field can improve the
$\sigma$ condensate, which is of course the chiral magnetic
catalysis effect again.

\begin{figure}

\begin{center}
\includegraphics[width=7cm]{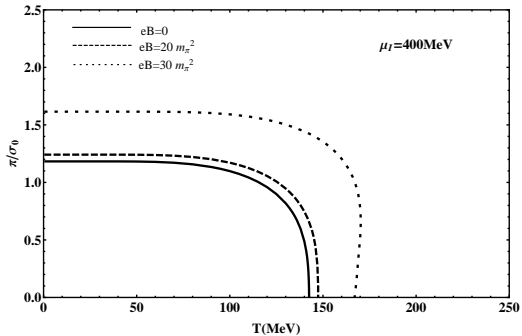} \\
\end{center}

\caption{The Pion condensate $\pi$, shown as a function of the
temperature with respect to various magnetic field magnitude at
fixed isospin chemical potential $\mu_I=400MeV$.} \label{pt400}
\end{figure}

The influence of the external magnetic field on the pion
condensation is different for different $\mu_I$. Fig. 5 shows the
change tendency of the pion condensation with  $\mu_I=400MeV$.

For fixed $eB$, the $\pi$ condensate decreases monotonously with
increasing temperature, that coincides with the case when
$\mu_I=100MeV$. This implies that the temperature always depress
the pion condensation. Taking the magnetic field into
consideration, the greater the $eB$, the wider range of
temperature domain of the $\pi\neq0$ and the larger the critical
temperature for $\pi$ superfluidity transition,reflecting that
$\pi$ condensate is promoted with the strong magnetic field for
$\mu_I=400MeV$. Comparing the varying behaviors of the pion
condensation for $\mu_I=100MeV$ with that for $\mu_I=400MeV$, we
prove again the influence of the external magnetic field on the
pion condensation is not simple promotion or suppression. The
effect of the external magnetic field on the pion superfluid is
related to the value of the isospin chemical potential.

\subsection {the pion superfluidity phase diagram}

\begin{figure}
\begin{center}
\includegraphics[width=7cm]{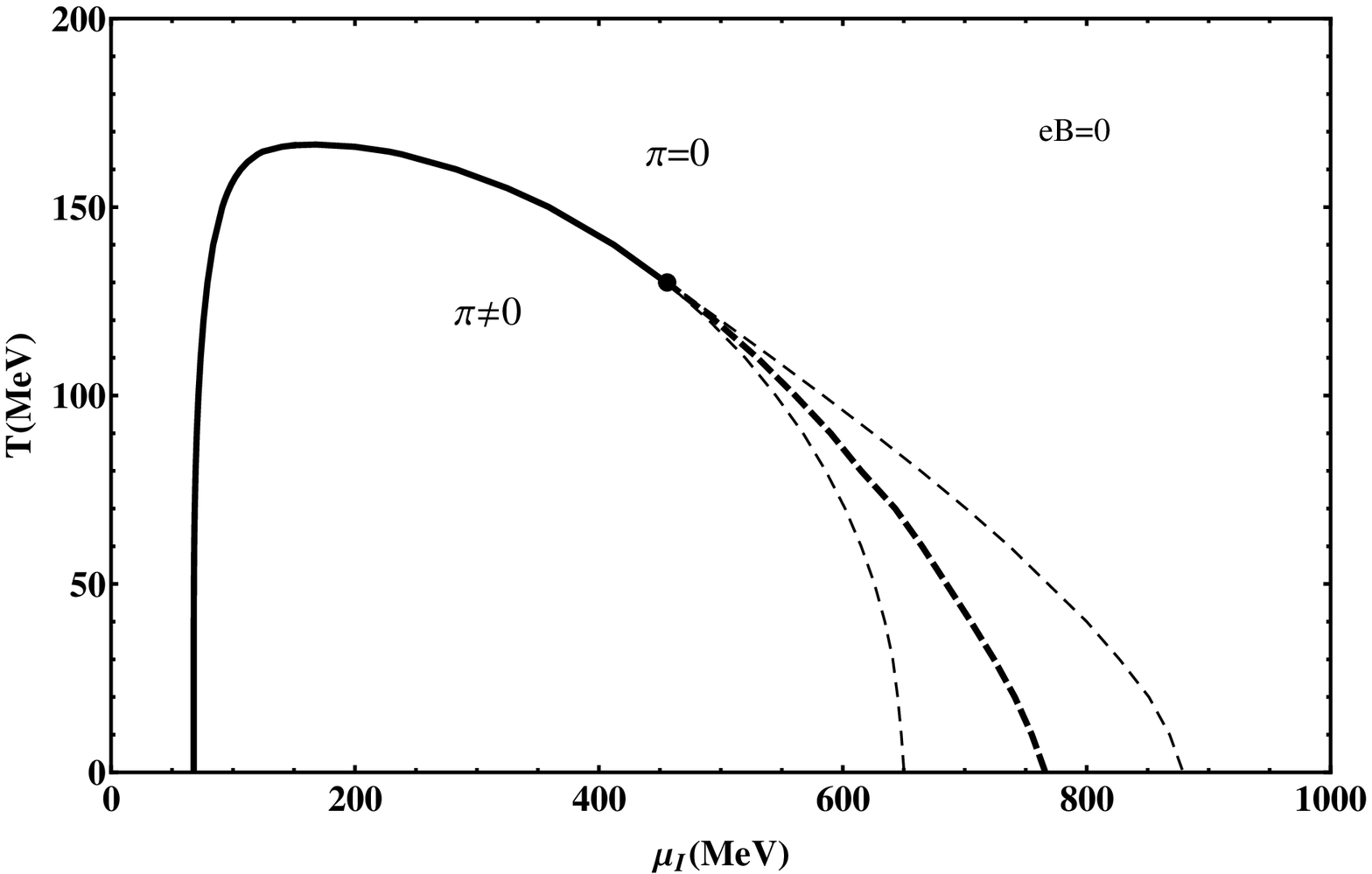} \\
\includegraphics[width=7cm]{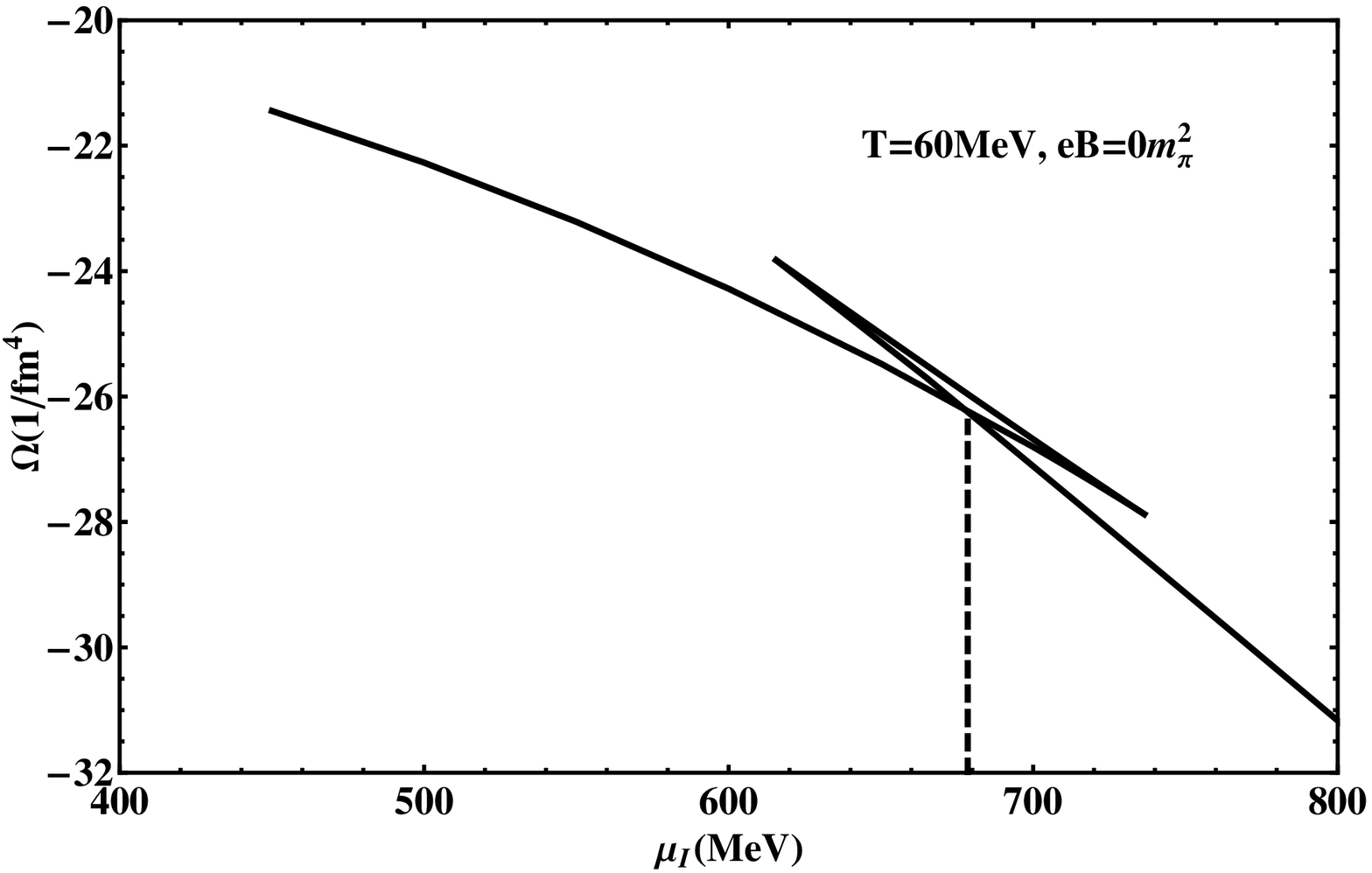} \\
\end{center}
\caption{{\em Upper panel:} The Pion superfluidity phase diagram
in $T-\mu_I$ plane at $eB=0$. {\em Lower panel:} The thermodynamic
potential as a function of $\mu_I$ at $eB=0$ and $T=60MeV$.}
\label{duiteb}
\end{figure}

From the change in the $\pi$ and $\sigma$ order parameter as a
function of $T, eB,$ and $\mu_I$, we can  calculate both the pion
superfluidity and the chiral phase transition phase diagram. We
confine our analysis to the pion phase diagram in this paper. Here
we define the transition happens when the $\pi$ condensation goes
to zero, namely the pion superfluidity phase transform to the
normal phase.

In the upper panel of Fig. \ref{duiteb} we plot the pion
superfluidity phase diagram in $T-\mu_I$ plane at $eB=0$. The
region inside of the curve with $\pi \neq 0$ is superfluid phase,
and the region outside of the curve with $\pi = 0$ is normal
phase. When $\mu_I$ is small, the phase transition is second
order, and then it becomes first order at large $\mu_I$. The point
connects the first and the second order phase transition lines is
the so-called tricritical point. At $eB=0$, this point is
approximately at $(T,\mu)=(456,130)$. The two thin dashed lines
are the metastable lines of the first order phase transition,
which can be observed from the upper panel of Fig. 1. The isospin
chemical potential on the higher dashed line is the maximal
$\mu_I$ when the pion condensation exists, and on the lower dashed
line the $\mu_I$ is the point where the pion condensate
disappears. The thick dashed line in the middle is the so-called
Maxwell line. The points on the Maxwell line meet the condition of
phase equilibrium of the first order phase transition with
$p_1=p_2, \mu_{I1}=\mu_{I2}$ and $T_1=T_2$.

The lower panel of Fig. \ref{duiteb} clearly shows how we
determine this Maxwell line. We plot the behavior of the
thermodynamic potential $\Omega(=-P)$ with varying $\mu_I$, for
example, for $T=60 MeV$ and $eB=0$. From the cross point on the
figure we obtain the value of $\mu_I$ at this temperature. By
changing the temperature and doing the same calculation, we then
obtain the middle dashed line on the upper panel of Fig. 6.

\begin{figure}
\begin{center}
\includegraphics[width=7cm]{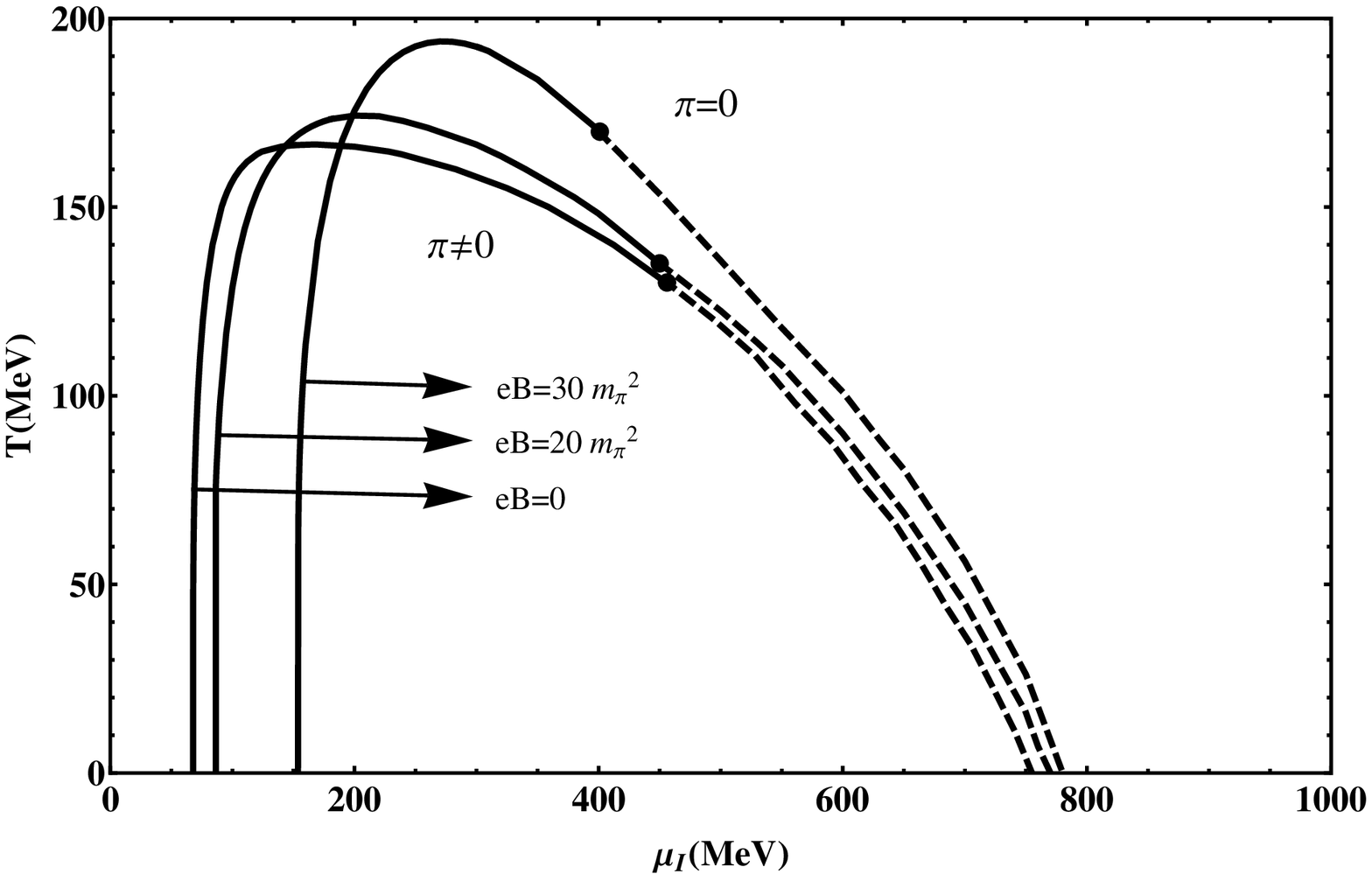} \\
\includegraphics[width=7cm]{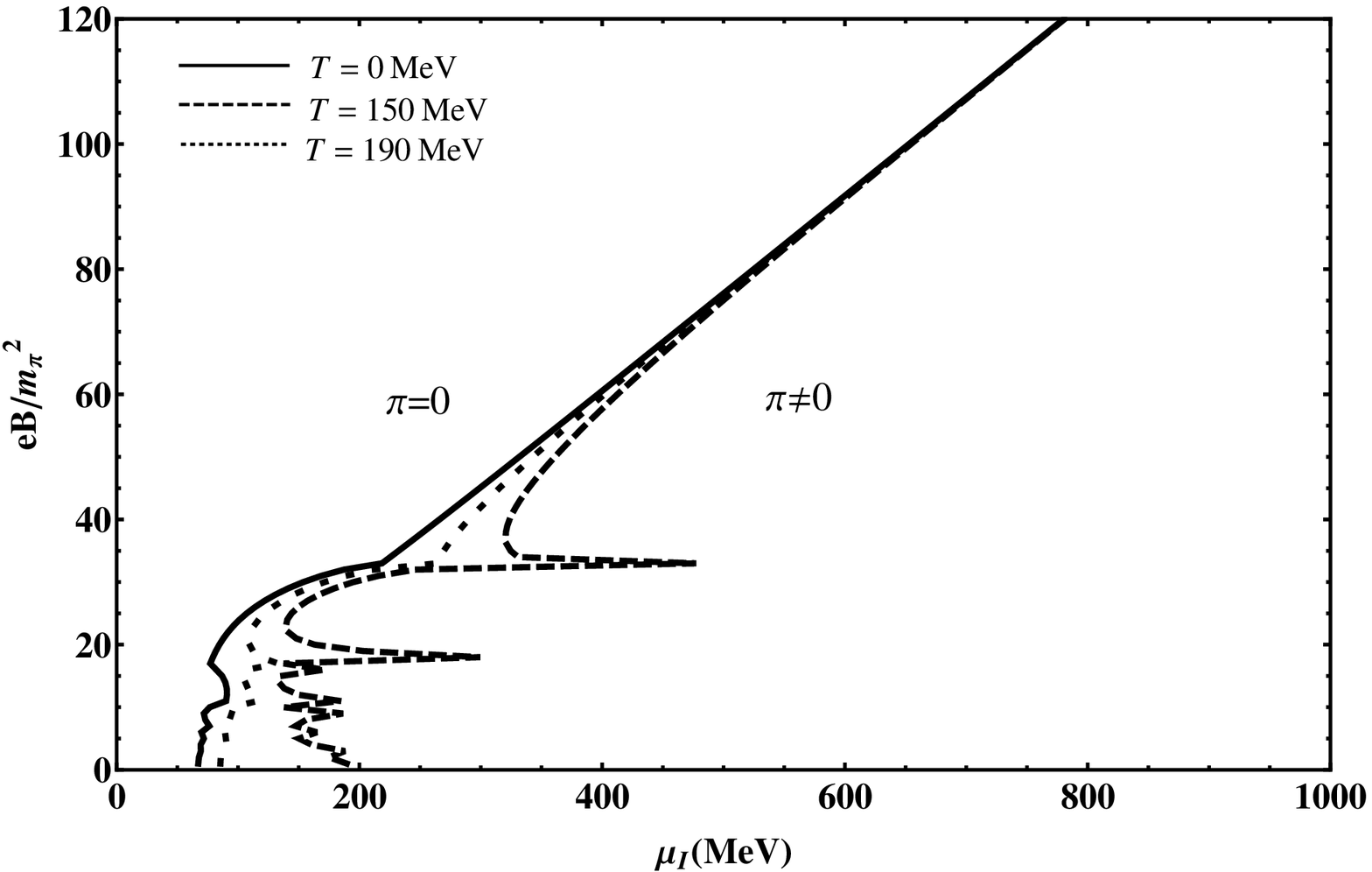} \\
\includegraphics[width=7cm]{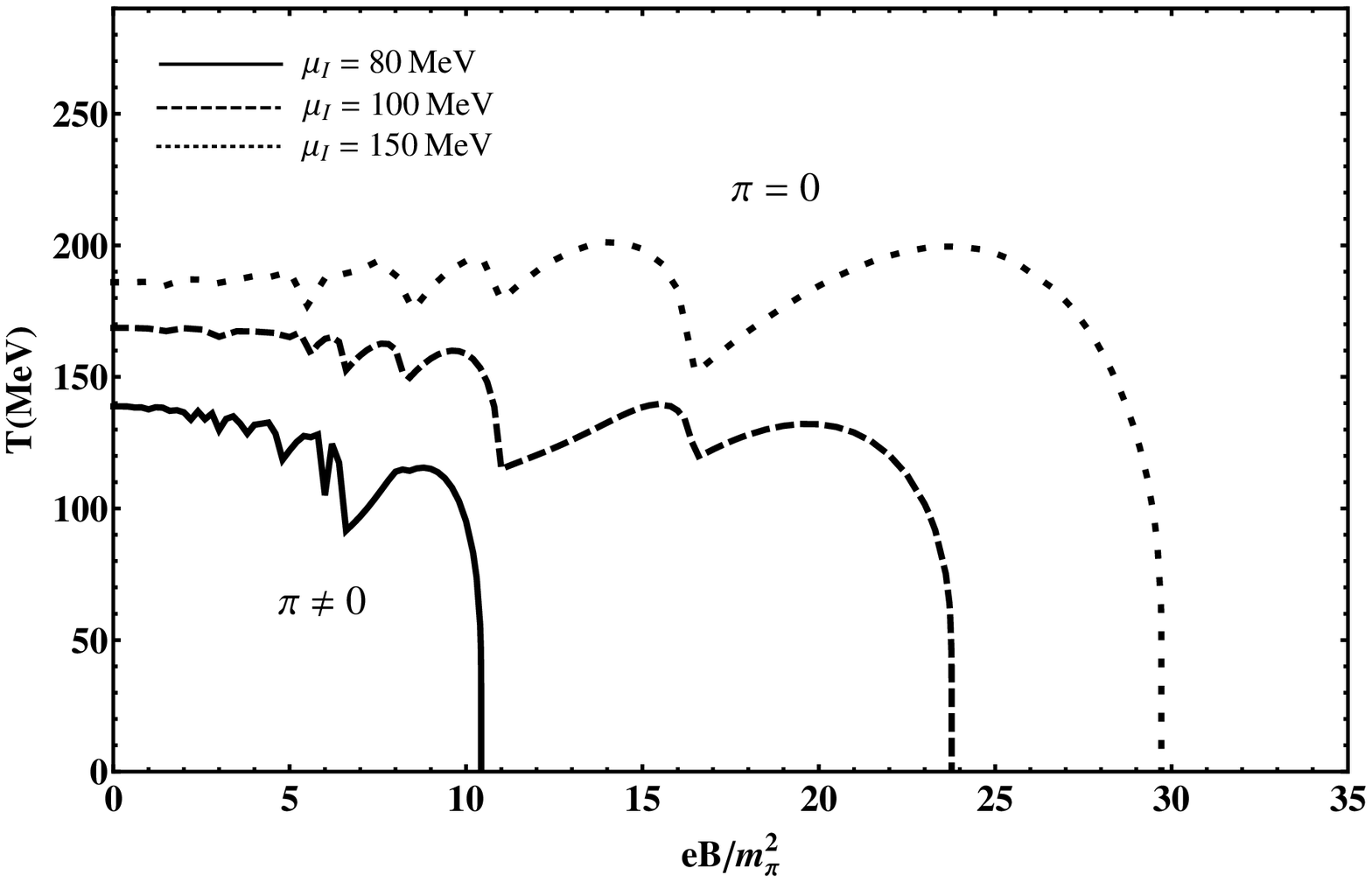} \\
\end{center}
\caption{{\em Upper panel:} The pion superfluidity phase diagram
in $T-\mu_I$ plane. {\em Middle panel:} The pion superfluidity
phase diagram in $eB-\mu_I$ plane. {\em Lower panel:} The pion
superfluidity phase diagram in $T-eB$ plane.} \label{tebui100}
\end{figure}

Fig. \ref{tebui100} show the pion superfluidity phase diagram of
our model including the strong magnetic field effect. In the upper
panel, we show the pion superfluidity phase diagram in $T-\mu_I$
plane at different magnetic field, $eB=0$, $20m_\pi^2$, and
$30m_\pi^2$, respectively. The regions outside of the curves are
in normal phase. The regions inside the curves with lower $T$ are
the $\pi$ superfluidity phases.

For $eB=0$ case, when $\mu_I$ is larger than $\mu_I^c$ (about
$m_\pi/2$), the $\pi$ superfluidity phase forms. The pion
superfluidity vanishes with increasing $T$. When the external
magnetic fields are included, such as $eB=20 m_\pi^2$ and $30
m_\pi^2$, the regions of $\pi\neq0$ shrink as the $eB$ increases
for $\mu_I<200MeV$, which can be verified by the upper panel in
Fig.4. When $\mu_I$ is very large, for $\mu_I>200MeV$, the regions
of $\pi\neq0$ become larger as the $eB$ increases, this phenomenon
is coincide with the lower panel in Fig.4. All in all, the region
of $\pi\neq 0$ is nonmonotonously affected by the external $eB$.
From our calculation, the tricritical point moves to the space
with smaller $\mu_I$ and higher $T$ when the external magnetic
field becomes stronger.

The middle panel shows us the pion superfluidity phase diagram in
$eB-\mu_I$ plane at different values of temperature, $T=0$,
$150MeV$, and $190MeV$, respectively. The regions in the left-side
of the curves are the normal phases and the right-side of the
curves with higher $\mu_I$ are the pion superfluidity phase. The
regions of $\pi\neq0$ decrease with increasing $T$ because the
temperature will suppress the formation of the $\pi$
superfluidity.

In the upper panel, we can roughly conclude that the critical
$\mu_I^c$ where the pion superfluidity occurs increases as the
$eB$ increases. If we investigate it in detail, shown in the
middle panel, the $\mu_I^c$ actually does not rise monotonously
but oscillatorily as the $eB$ increases when the magnetic field is
not so strong. This oscillation is related to the splitting quark
energy level in the surrounding of the magnetic field. For
$eB>40m_\pi^2$, the $\pi$ superfluid phase boundary becomes
monotonous and smooth with the increasing $\mu_I$. This is because
the Lower Landau Level (LLL) chiefly contributes to the behaviors
of phase transition boundaries for large $eB$.

In the lower panel we show the pion superfluidity phase diagram in
$T-eB$ plane for different values of the isospin chemical
potential, $\mu_I=80MeV$, $100MeV$, and $150MeV$, respectively.
The regions inside the curves with low $eB$ and low $T$ are the
$\pi$ superfluidity phases. The larger the $\mu_I$, the wider the
region of $\pi\neq0$. This means that the $\pi$ superfluid is
enhanced with increasing $\mu_I$. Here we still obtain the
oscillated phase boundary for the pion superfluid phase and this
results to a interesting phenomenon with the increasing $eB$.
Taking $\mu_I=150MeV$ and $T=185MeV$ for an example, when
increasing the $eB$ the pion superfluidity appears occasionally.
This phenomenon of the superfluid phase carry on alternatively.
The oscillatory behavior of the pion superfluid phase boundary is
also a typical subsequence of the magnetic field effect. This
result is similar with the conclusion of the middle panel in this
figure.

\section {Conclusion and discussion}
\label{s4}
The effect of the external magnetic field on the pion condensate
and chiral condensate is investigated in the two flavor NJL model
at finite temperature and isospin potential. The chiral and pion
condensation are calculated in the surrounding of the various
magnetic field. The oscillating behavior of the chiral and pion
condensation, related to the so-called Alfven-de Haas oscillation,
is shown by the numerical result. The phase boundary of the pion
superfluid are also investigated with the consideration of the
magnetic field. An oscillatory phase boundary is found with
increasing the magnetic field. The first order phase transition in
the high isospin chemical potential region still exists when the
magnetic field effect is taken into account. With the magnetic
field considered, the tricritical point on the phase diagram of
the $T-\mu_I$ plane moves to the high $T$ and low $\mu_I$ region.
Because of the discrete of Landau Level, the influence of the
external magnetic field on the pion condensation and the phase
boundary is not simple promotion or suppression, which is
different from the effect of the temperature and the isospin
chemical potential. The $\sigma$ condensate is enhanced with
increasing magnetic field, which is supported by the chiral
magnetic catalysis effect. However, because of the oscillation of
the chiral condensate, there exist some magnetic field region
where the chiral condensation would decrease with the increasing
$eB$.

{\bf Acknowledgement:} This work was supported in part by NSFC
(grants 11375070).

\end{document}